\documentstyle[12pt]{article}
\def\singlespace {\smallskipamount=3.75pt plus1pt minus1pt
                  \medskipamount=7.5pt plus2pt minus2pt
                  \bigskipamount=15pt plus4pt minus4pt
                  \normalbaselineskip=15pt plus0pt minus0pt
                  \normallineskip=1pt
                  \normallineskiplimit=0pt
                  \jot=3.75pt
                  {\def\smallskip {\vskip\smallskipamount}}
                  {\def\medskip   {\vskip\medskipamount}}
                  {\def\bigskip   {\vskip\bigskipamount}}
                  {\setbox\strutbox=\hbox{\vrule
                    height10.5pt depth4.5pt width 0pt}}
                  \parskip 7.5pt
                  \normalbaselines}
                  \def\middlespace {\smallskipamount=5.825pt plus1.5pt minus1.5pt
                  \medskipamount=11.25pt plus3pt minus3pt
                  \bigskipamount=22.5pt plus6pt minus6pt
                  \normalbaselineskip=22.5pt plus0pt minus0pt
                  \normallineskip=1pt
                  \normallineskiplimit=0pt
                  \jot=5.825pt
                  {\def\smallskip {\vskip\smallskipamount}}
                  {\def\medskip   {\vskip\medskipamount}}
                  {\def\bigskip   {\vskip\bigskipamount}}
                  {\setbox\strutbox=\hbox{\vrule
                    height15.75pt depth6.75pt width 0pt}}
                  \parskip 7.25pt
                  \normalbaselines}
\def\dblspc {\smallskipamount=7.5pt plus2pt minus2pt
                  \medskipamount=15pt plus4pt minus4pt
                  \bigskipamount=30pt plus8pt minus8pt
                  \normalbaselineskip=30pt plus0pt minus0pt
                  \normallineskip=2pt
                  \normallineskiplimit=0pt
                  \jot=7.5pt
                  {\def\smallskip {\vskip\smallskipamount}}
                  {\def\medskip   {\vskip\medskipamount}}
                  {\def\bigskip   {\vskip\bigskipamount}}
                  {\setbox\strutbox=\hbox{\vrule
                    height21.0pt depth9.0pt width 0pt}}
                  \parskip 15.0pt
                  \normalbaselines}
\def\un{\underline}
\def\la{\lambda}
\def\dis{\displaystyle}

\def\al{\alpha }
\def\gm{\gamma}
\def\l{\left}
\def\r{\right}
\def\o{\omega}
\def\f{\frac}
\def\p{\partial}
\def\nb{\nabla}
\def\eps{\epsilon}

\def\be{\begin{equation}}
\def\ee{\end{equation}}
\def\barr{\begin{array}}
\def\earr{\end{array}}
\begin{document}
\dblspc
\begin{center}
{\Large{\bf Gravitational Wave Induced Rotation\\[8pt]
of the Plane of Polarization of Pulsar Signals}}\\[20pt]

A.R. Prasanna and S.~Mohanty\\
Physical Research Laboratory\\
Navrangpura, Ahmedabad 380 009\\
India\\[20pt]
\end{center}

\begin{center}
{\bf Abstract}
\end{center}

 We derive in this an expression for the rotation of
plane of polarization, of an electromagnetic wave, induced by the
field of a gravitational wave propagating along the same direction
$\approx \f{G\mu d^2\Omega^4}{3\o}$, $\o$ and $\Omega$ being their
respective frequencies using the geometrical optics limit of the
wave equation for fields. Estimating the effect for the case of
pulses from  binary pulsars, one finds it  too small to
be
observable, there could be other
sources like inspiralling binary or an asymmetric neutron star
where the effect could be in the observable region.

04.30.Nk , 42.15.i, 97.60Gb, 97.80d

\newpage

The existence of gravitational waves is although confirmed from
the binary pulsar data, attempts to detect directly have been
pursued for over three decades with all the available technology.
In this letter we pose a question whether one could detect these
waves through their interaction with electromagnetic waves. In
fact, both electromagnetism and gravitation being long range
interactions, described through respective fields, it is possible
that there exist an interaction between the two and some imprint
of gravitational waves gets etched on electromagnetic waves
through space time curvature. As it is far more easier to detect
electromagnetic waves, we propose that using wave equation for
electromagnetic field components on curved space time, one could get a
dispersion relation in the geometrical optics limit that shows a
formal rotation of the plane of polarization of the
electromagnetic wave. The magnitude of the angle of rotation $\l(
\sim \f{GM \Omega^4}{kC^5}\r)$, which depends on the fourth power of
the gravitational wave frequency could become detectable through
pulsar radiation (Han et al 1998),
 for systems with very high frequency $\Omega$.\\

It has been shown  (Mohanty and Prasanna, 1998) that the wave
equation for the electromagnetic field tensor in curved spacetime
involves a coupling with the Riemann as well as Ricci curvatures
of the background spacetime, and is given by
 \be \nb^\mu
\nb_\mu F_{\al\beta} + R_{\al\beta\mu\nu} F^{\mu\nu} +
R^\mu_{\;\;\al} F_{\beta\mu} - R^\mu_{\;\;\beta} F_{\al\mu} = 0 .
\ee Considering the space time outside material distribution
$R_{\mu\nu} = 0$, but having non-zero curvature produced by plane
gravitational waves we look for a solution of (1) in the
geometrical optics limit (eikonal approximation) with the ansatz
\be \barr{lll} F_{\al\beta}& =& e^{is/\eps} f_{\al\beta}\\[12pt]
f_{\al\beta}&=& {\sum_n}\l( \f{\eps}{i} \r)^n f_{n\; \al\beta} \earr \ee
and \be g_{\al\beta} = \eta_{\al\beta} + \hat{g}_{\al\beta} \ee
\be \barr{lll}
\hat{g}_{\al\beta}&=& e^{is/\eps} \gm_{\al\beta}\\[12pt]
\gm_{\al\beta}&=& {\sum_n} \l( \f{\eps}{i} \r)^n \gm_{n\; \al\beta}
\earr
\ee (Ehlers and Prasanna, 1996), wherein the phase functions $s$
(electromagnetic) and $S$ (gravitational) are rapidly varying as
compared to their respective amplitudes and the smallness
parameter $\eps = \f{\la}{L}$, is the ratio of wavelength to the
background scale length. We shall first express eq. (1) in
geodesic normal coordinates to get \be \p^\mu \p_\mu \;
F_{\al\beta} + \f{1}{2} \l( g_{\beta\mu , \al\nu} + g_{\al\nu ,
\beta\mu} - g_{\al\mu , \beta\nu} - g_{\beta\nu , \al\mu}\r)
F^{\mu\nu} = 0 \ee Using (2)-(4) in (5) and denoting the wave
vectors \be \p_\mu s = k_\mu \mbox{ and } \p_\mu S = \ell_\mu \ee
one gets \be \barr{ll} &e^{is/\eps} \l\{ - \f{1}{\eps^2} k^\mu
k_\mu f_{\al\beta} + \f{i}{\eps} \l( \p^\mu k_\mu + 2k_\mu \p^\mu
\r) f_{\al\beta} +
\p^\mu \p_\mu f_{\al\beta} \r\}\\[12pt]
&+ \f{1}{2} e^{iS/\eps} \l\{ - \f{1}{\eps^2} \l( \ell_\nu
\ell_\al \gm_{\beta\mu} + \ell_\beta \ell_\mu \gm_{\nu\al} -
\ell_\beta \ell_\nu \gm_{\al\mu} - \ell_\al \ell_\mu
\gm_{\beta\nu} \r) \right.\\[12pt]
&+ \f{i}{\eps} \l[ \l( \p_\nu \ell_\al \p_\nu + \ell_\nu \p_\al
\r) \gm_{\beta\mu} + \l( \p_\mu \ell_\beta + \ell_\beta \p_\mu +
\ell_\mu \p_\beta \r) \gm_{\al\nu}\right. \\[12pt]
&\left. - \l( \p_\nu \ell_\beta + \ell_\beta \p_\nu \p_\beta \r)
\gm_{\al\mu} - \l( \p_\nu \ell_\al + \ell_\al \p_\mu + \ell_\mu
\p_\al \r) \gm_{\beta\nu} \r]\\[12pt]
&\left. + \l( \p_\nu \p_\al \gm_{\beta\mu} + \ell_\beta \p_\mu
\gm_{\nu\al} - \p_\nu \p_\beta \gm_{\al\mu} - \p_\mu \p_\al
\gm_{\nu\beta} \r) \r\} e^{is/\eps} f^{\mu\nu} = 0 \earr \ee
Collecting the zeroth order terms in the amplitudes the equation
reduces to \be k^\mu k_\mu f_{0\al\beta} + \f{1}{2} \l\{ \ell_\nu
\ell_\al \gm_{0\beta\mu} + \ell_\beta \ell_\mu \gm_{\nu\al} -
\ell_\beta \ell_\nu \gm_{0 \al\mu} - \ell_\al \ell_\mu \gm_{0
\beta\nu} \r\} f_0^{\mu\nu} = 0 \ee Assuming the curvature to be
due to plane monochromatic gravitational wave propagating in the
z-direction, the space time metric would be given by \be dS^2 =
dt^2 - \l( 1 - h_{xx} \r) dx^2 - \l( 1 + h_{xx} \r) dy^2 + 2h_{xy}
dxdy-dz^2 \ee
 Further, considering the electromagnetic wave also
to be along the z-direction, one has the folowing wave vectors \be
\barr{lll}
\ell_\al&=& \l( 0, 0, \ell , \Omega \r)\\[12pt]
k_\al&=& \l( 0, 0, k, \o \r) \earr \ee for the gravitational and
electromagnetic waves respectively, and  equation (7) takes the
form: \be k^\mu k_\mu f_{oi} + e^{iS/\eps} \Omega \l( \ell_\nu
\gm_{ij} f^{j\nu} \r) = 0 \ee which may be written for the
electric fields $f^{01}$ and $f^{02}$ \be \l( \barr{cc} \l( \o^2 -
k^2 \r)+ \Omega^2 \gm_{11}
e^{iS/\eps}&\Omega^2\gm_{12} e^{iS/\eps}\\[12pt]
\Omega^2\gm_{21} e^{iS/\eps}&\l( \o^2-k^2\r) +\Omega^2 \gm_{22}
e^{iS/\eps} \earr \r) \l( \barr{c} f^{01}\\[12pt] f^{02} \earr
\r) = 0
\ee
The condition for the non-trivial solution for $f^{01}$,
$f^{02}$ yields the dispersion relation
\[
\l( \o^2-k^2\r)^2 - \Omega^4 \l( \gm_{11}^2 + \gm^2_{12} \r)
e^{2iS/\eps} = 0
\]
or \be k^2= \o^2 \pm \Omega^2 \sqrt{\gm^2_{11} + \gm^2_{12}}
~~e^{iS/\eps} \ee It is well known that for the field of the weak
gravitational wave, from the point masses in a Keplerian orbit
(binary system), the metric potentials are given by (Peters and
Mathews, 1963) \be \barr{lll} h_{xx}&=& -\f{G\Omega^2}{3 r} \l(
Q_{xx} - Q_{yy} \r)
e^{i\Omega (t-z )} = - h_{yy}\\[12pt]
h_{xy}&=& - \f{2G\Omega^2}{3r } Q_{xy} e^{i\Omega (t-z )} \earr
\ee with \be \barr{lll}
Q_{xx}& = & \mu d^2 \cos^2\psi\\[12pt]
Q_{yy}&=& \mu d^2\sin^2\psi\\[12pt]
Q_{xy}&=& \mu d^2 \sin \psi \cos\psi \earr \ee $\mu =
\f{M_1M_2}{M_1+M_2}$ being the reduced mass of $M_1$, $M_2$,the
two pulsar masses and $d = a \f{\l( 1 - e^2\r)}{\l( 1 + e
\cos\psi\r)}$ while the angular velocity \be \dot{\psi} = \f{\l[ G
\l( M_1 + M_2 \r) a \l( 1 - e \r) \r]^{1/2}}{d} \ee As we have
$\gm_{11} = h_{xx} e^{iS/\eps}$, $\gm_{12} = h_{xy} e^{-iS/\eps}$
, using (14) and (15) in (13), one finds \be k^2 = \o^2 \l( 1
\pm \f{G\mu d^2\Omega^4}{3\o^2 r} e^{i\Omega (t-z )} \r) \ee Since
the gravitational waves propagate with the same phase velocity as
electromagnetic waves, one gets the dispersion relation for
electromagnetic waves to be \be k_\pm = \o \pm \f{G\mu
d^2\Omega^4}{3\o r} \ee Thus the electromagnetic pulse passing
through the field of gravitational waves will have its plane of
polarisation rotated through an angle \be \Delta \phi =
{\dis{\int^{r_{max}}}_{r_{min}}} dr \l( k_+ - k_-\r) = \f{G\mu
d^2\Omega^4}{3\o} \ell n \l( \f{r_{max}}{r_{min}} \r) \ee Consider
the follwing  examples:
\begin{itemize}
\item[(a)] For the Hulse-Taylor binary 1913+16, the values of the
relevant parameters are $\mu = 0.7 M_\odot$, $d  \sim 2 $ secs,
orbital period = $P = \f{2\pi}{\Omega}$ = 2.7906.98 sec (Weisberg
and Taylor 1984) . Considering the pulsar beam of frequency $\o
\approx $ 400 MHz, the optical rotation over a distance of $\sim
10^2 kpc$ is $\Delta
\phi \approx 10^{-28} rad$, which is very small.\\
\item[(b)] On the other hand, if one considers the gravitational waves from an
 inspiralling
binary (Nakamura et al 1998) with a pulsar with $d\approx 100 km$,
between the two compact objects, the orbital frequency $\Omega =
\l[ \f{G \l( M_1 + M_2 \r)}{d^3} \r]^{1/4}$ $\sim$ $5\times
10^{-3}$ sec. for $M_1$ $\sim$ $M_2$ $\sim$ 1.4 $M_\odot$. For
this system, the optical rotation of a 400 MHz radio signal over a
distance of $\sim 100 Mpc$ turns out to be $\Delta \phi \approx
1.9 \times 10^{-10} rad$.
\item[(c)] Finally considering the
gravitational waves from an assymetric pulsar (Thorne 1997) with a
non-zero quadrupole moment parametrized by $\eta GMR^2$, the
emitted gravitational wave will have the amplitude $h \approx
\f{\eta GMR^2\Omega^2}{r}$. Using this for the metric potential
$h_{xx} = h_{yy} \approx h$, the dispersion relation for the
electromagnetic signal would be \be k_\pm = \o \l[ 1 \pm \f{\eta
GMR^2\Omega^4}{\sqrt{2} r\o^2} \r] \ee
 and the corresponding optical rotation, for $M \sim 1.4
M_\odot$, $R \sim 10$ km, $\Omega = \f{2\pi}{10^{-3}} s$, over a
distance of 100 $Mp_c$ is \be \Delta \phi \approx 2.8 \eta \times
10^{-8} \mbox{ radians} \ee
 If
$\eta$ is not too small, it may be possible to observe the
signature of gravitational waves from such a pulsar.
\end{itemize}

In conclusion, it may be noted that the electromagnetic radiation
emitted from a region which is also emitting sufficiently strong
gravitational radiation, would suffer gravity induced
birefringence which over a long distance can materialise as a
rotation of the plane of polarization of the electromagnetic
signal, thus establishing the presence of gravitational waves.

\un{References}
\begin{enumerate}
\item Ehlers J. and Prasanna A.R. (1996) {\it Class. \& Quan.
Grav.}, {\bf 13}, 2231.
\item Han J.L, Manchester R.N, Xu R.X , Qiao C.J,(1998) MNRAS {\bf
300},373.
\item Mohanty S. and Prasanna A.R. (1998) {\it Nucl. Phys.} {\bf
B526},  501.
\item Peters P.C. and Mathews J. (1963) {\it Phys. Rev.} {\bf
131}, 435.
\item Weisberg J.M and Taylor, J. H (1984), Phys Rev Lett
{\bf 52},1348.
\item Nakamura T, Sasaki M,Tanaka T,Thorne K.S, (1997) ApJ
{\bf487},L139.
\item Thorne K.S, (1997) Archive gr-qc /9704042
\end{enumerate}

\end{document}